\documentclass[preprint,authoryear,12pt]{elsarticle}
\usepackage{epsfig}
\usepackage{amssymb}
\usepackage[ps2pdf,%
a4paper=true,%
breaklinks=true,%
colorlinks=true,%
pdfauthor={First Author et al.},%
pdftitle={Template for manuscripts in Advances in Space Research}%
]{hyperref}
\journal{Advances in Space Research}

\begin{document}
\begin{frontmatter}
\title{The Contribution of Millisecond Pulsars\\ to the Galactic Cosmic-Ray Lepton Spectrum}
\author[NWU]{Christo Venter\fnref{email}}
\address[NWU]{Centre for Space Research, North-West University, Potchefstroom Campus, 2520 Potchefstroom, South Africa}
\fntext[email]{Email: Christo.Venter@nwu.ac.za}

\author[NWU]{Andreas Kopp\fnref{footnote1}}
\fntext[footnote1]{On leave from Institut f\"ur Experimentelle und Angewandte Physik, Christian-Albrechts-Universit\"at zu Kiel, Leibnizstrasse 11, 24118, Kiel, Germany}

\author[NASA]{Alice K Harding}
\address[NASA]{Astrophysics Science Division, NASA Goddard Space Flight Center, Greenbelt, MD 20771, USA}

\author[Hope]{Peter L Gonthier}
\address[Hope]{Hope College, Department of Physics, Holland, MI, USA}

\author[NWU]{Ingo B\"usching}

\begin{abstract}
Pulsars are believed to be sources of relativistic electrons and positrons. The abundance of detections of $\gamma$-ray millisecond pulsars by \textit{Fermi} Large Area Telescope coupled with their light curve characteristics that imply copious pair production in their magnetospheres, motivated us to investigate this old pulsar population as a source of Galactic electrons and positrons and their contribution to the enhancement in cosmic-ray positron flux at GeV energies. We use a population synthesis code to predict the source properties (number, position, and power) of the present-day Galactic millisecond pulsars, taking into account the latest \textit{Fermi} and radio observations to calibrate the model output. Next, we simulate pair cascade spectra from these pulsars using a model that invokes an offset-dipole magnetic field. We assume free escape of the pairs from the pulsar environment. We then compute the cumulative spectrum of transported electrons and positrons at Earth, following their diffusion and energy losses as they propagate through the Galaxy. Our results indicate that the predicted particle flux increases for non-zero offsets of the magnetic polar caps. Comparing our predicted local interstellar spectrum and positron fraction to measurements by \textit{AMS$-$02}, \textit{PAMELA}, and \textit{Fermi}, we find that millisecond pulsars are only modest contributors at a few tens of GeV, after which this leptonic spectral component cuts off. The positron fraction is therefore only slightly enhanced above 10~GeV relative to a background flux model. This implies that alternative sources such as young, nearby pulsars and supernova remnants should contribute additional primary positrons within the astrophysical scenario.
\end{abstract}

\end{frontmatter}

\parindent=0.5 cm
\section{Introduction}
\label{sec:intro}
\textit{PAMELA} was the first experiment to provide firm evidence of a rising positron fraction (PF), defined as $\phi(e^+)/[\phi(e^+) + \phi(e^-)]$, of the local (leptonic) interstellar spectrum above $\sim10$~GeV \citep{Adriani09}. The \textit{Fermi} Large Area Telescope (LAT) has confirmed this result \citep{Ackermann12}, and \textit{PAMELA} has recently extended the measurement range up to 300 GeV \citep{Adriani13}. \textit{AMS$-$02} has now provided a PF measurement in the range 0.5 $-$ 350 GeV \citep{Aguilar13} for 18 months of data, and improved spectra for 30 months of data \citep{Aguilar14,Accardo14}, extending the PF up to 500 GeV and indicating a levelling off of this fraction with energy, as well as finding the PF to be consistent with isotropy. 

Positrons are created during inelastic collisions involving cosmic-ray nuclei and intergalactic hydrogen, which produce charged pions that in turn decay into positrons, electrons, and neutrinos. The fraction of such a \textit{secondary} positron component with respect to the total leptonic cosmic-ray spectrum is expected to smoothly decrease\footnote{This, however, depends on model assumptions, i.e., a concave electron spectrum may lead to a rising PF.} with energy \citep[e.g.,][]{Moskalenko98}. However, the \textit{AMS$-$02} electron spectrum is softer than the positron flux in the range $20 - 200$~GeV \citep{Aguilar14}, and the measured PF rises with energy, pointing to nearby sources of \textit{primary} positrons. Moreover, the rising PF can be ascribed to a hardening of the positron spectrum (up to 200~GeV, after which it softens with energy), and not a softening in electron spectrum above 10~GeV. 

The purported additional source of primary positrons may be either of astrophysical or dark matter annihilation origin \citep[e.g.,][]{Fan10,Porter11}. The first case may include supernovae \citep[e.g.,][]{Blasi09,D10}, microquasar jets \citep{Gupta14} pulsar wind nebulae \citep[e.g.,][]{Serpico12}, mature pulsars \citep{Zhang01}, and millisecond pulsars \citep[MSPs;][]{Kisaka12}. The detection of a potential anisotropy in lepton flux may be an important discriminator between dark matter vs.\ pulsar (or supernova remnant) scenarios \citep{Accardo14}, although this may depend on the number of individual sources that collectively contribute to the cosmic-ray lepton spectrum at Earth. 

Alternatively, there have been several attempts to modify the standard Galactic cosmic-ray transport models in order to explain the observed rise in PF with energy purely by secondary positrons originating in the interstellar medium. \citet{Shaviv09} demonstrated that an inhomogeneous distribution of SNRs, such as a strong concentration in the Galactic spiral arms, may explain the PF shape. \citet{Moskalenko13} pointed out that the concave shape of the primary electron spectrum of \citet{Shaviv09} introduces an arguably artificial rise in the PF. \citet{Cowsik10} put forward a model that assumes a significant fraction of the boron below 10~GeV is generated through spallation of cosmic-ray nuclei in small regions around the sources; GeV positrons would then almost exclusively be generated through cosmic-ray interactions in the interstellar medium. \citet{Moskalenko13} noted that such sources should be observable as very bright GeV $\gamma$-ray sources with soft spectra, while the diffuse emission would be significantly dimmer than observed. This scenario is also at odds with current estimates of the supernova birth rate. \citet{Blum13} found an upper bound to the positron flux by neglecting energy losses, arguing that the flattening of the PF seen by \textit{AMS-02} around several hundred GeV is consistent with a purely secondary origin for the positrons. \citet{Moskalenko13} noted that their arguments imply quite hard injection spectra for primary nuclei, in contradiction to $\gamma$-ray observations of SNRs. A very fast escape time for the positrons is furthermore implied, and if this is extrapolated to higher energies, it would lead to a large cosmic-ray anisotropy, which has not been observed.

The presence of $\gamma$-ray photons and intense magnetic fields in pulsar magnetospheres facilitate copious electron-positron pair production \citep{Sturrock71,DH82}, making pulsars prime candidate sources of primary Galactic leptons. We previously studied the contribution to the local electron spectrum by the nearby MSP PSR~J0437$-$4715 (assuming a pair-starved potential) as well as Geminga, and found that the latter may contribute significantly, depending on model parameters \citep{Buesching08}. \citet{Buesching08b} also noted that Geminga and PSR B0656+14 may be the dominant contributors to the local positron flux, and may be responsible for an anisotropy of up to a few percent in this flux component, depending on model parameters. \citet{Moskalenko13} however notes that to date the pulsar scenario lacks convincing calculations of the number of ejected particles and their spectrum. 

In this paper, we carefully assess the contribution of MSPs (excluding those found in the globular clusters) to the cosmic-ray lepton spectrum at Earth, given the large number of \textit{Fermi}-detected Galactic $\gamma$-ray MSPs \citep{Abdo13}, and the fact that their measured light curves imply abundant pair production even for this much older pulsar class \citep{Venter09}. We use a population synthesis code to predict the source properties (Section~\ref{sec:pop}) and a pair cascade code to find realistic injected pair spectra (Section~\ref{sec:pairs}). We next combine inverse Compton (IC) scattering of leptons on the local interstellar radiation field (ISRF; Section~\ref{sec:ISRF}) and their synchrotron radiation (SR) in the Galactic $B$-field (Section~\ref{sec:Bfield}) into an effective loss term (Section~\ref{sec:losses}), and use this together with a prescription for particle diffusion when solving a transport equation (Section~\ref{sec:transport}) to calculate the spectra at Earth (Section~\ref{sec:results}). We discuss our results Section~\ref{sec:concl}. More details will be provided in \citet{Venter15}.

\section{Present-Day Distribution of Galactic MSPs}
\label{sec:pop}
We implement the results of a new study by \citet{Gonthier15} of the population synthesis of radio and $\gamma$-ray MSPs that lead to the present-day distribution of MSPs. This distribution of MSPs is assumed to be an equilibrated one within the Galaxy. Its evolution has been described in Section~3 of \citet[][hereafter SGH]{Story07}, where the radial ($\rho$ in cylindrical coordinates) distribution was assumed to be that of the work of \citet{Pacz90}, with a radial scaling of 4.5~kpc and a scale height of 200~pc instead of 75~pc used in that work. In addition, we implement the supernova kick velocity model of \citet{Hobbs05} using a Maxwellian distribution with a width of 70~km\,s$^{-1}$ (resulting in an average of speed of 110 km\,s$^{-1}$). The Galaxy is seeded with MSPs treated as point particles with ages going back to the past 12~Gyr assuming a constant birth rate of $4.5\times10^{-4}$ MSPs per century as obtained in SGH. The MSPs are evolved in the Galactic potential from their birth location to the present time. Enough stars were evolved to obtain an equilibrium distribution of the present-day spatial distribution of MSPs.

We assume that MSPs are ``born'' on the spin-up line with initial period $P_0$ dependent on magnetic field, which we assume does not decay with time. We use a power-law distribution for the $B$-fields, similar to SGH. The simulation prefers a power-law distribution of periods $P(B_8)$, with $B_8 = B_0/(10^8$~G), and with an index $\alpha_{\rm B}$ (as in SGH). We achieve improved agreement with the new simulation with an index of $\alpha_{\rm B}=-1.3$ (as opposed to the value $\alpha_{\rm B}=-1$ used in SGH).

We assume a distribution of mass accretion birth lines from the Eddington critical mass accretion rate to about $10^{-3}$ of this value following the study by \citet{Lamb2005}. We parameterise the mass accretion rates with a line in the $\dot P-P$ diagram as was done in Eq.~(5) of SGH. 

Force-free electrodynamic solutions for the magnetosphere \citep{Spitkovsky06} led to the following description of the pulsar spin-down power $L_{\rm sd}$: 
\begin{equation}
L_{\rm sd} \sim \frac{2\,\mu^2\, \Omega^4}{3\,c^3}\left(1+\sin^2\alpha\right),\label{eq:Spit}
\end{equation}
where $\mu$ is the magnetic dipole moment, $\Omega$ is the rotational angular velocity, $c$ is the speed of light, and $\alpha$ is the magnetic inclination angle relative to the pulsar's rotational axis. This is similar to the results of \citet{Li12} for resistive magnetospheres and those of \citet{Contopoulos14} for force-free magnetospheres with current layers that depart from these ideal conditions. Equating $L_{\rm sd}$ to the rate of rotational energy loss, solving for $B_0$, and integrating this result over the age $t$ of the pulsar provides the present-day period $P$ from the pulsar's birth period $P_0$:
\begin{equation}
P^2 =P_0^2 +  \frac{ 4 \pi^2\,R^6}{3\,c^3 I } \left(1+\sin^2\alpha\right)\,B_0^2\, t.\label{eq:PresP}
\end{equation}
We assume $R\,=\,12\ {\rm km}$ and $M\,=\,1.6\ M_\odot$, where $M_\odot$ is the mass of the Sun. We use a moment of inertia of $I = 1.7 \times 10^{45}\  {\rm g\,cm^2}$ \citep{Pierbattista12}.

Our population synthesis model has been calibrated against \textit{Fermi} LAT data as well as those from a dozen radio surveys. Details, as well as comparison of simulated and observed radio and $\gamma$-ray pulsar properties (e.g., histograms of period, period derivative, surface magnetic field, characteristic age, and distance) will appear in our follow-up paper \citep{Venter15}, as well as in \citet{Gonthier15}.

\section{Electron-Positron Pair Spectra}
\label{sec:pairs}
We calculate the source spectra using a code that follows the development of a polar cap electron-positron pair cascade in the pulsar magnetosphere \citep[details of the calculation can be found in][]{Harding11}. The pair cascade is initiated by curvature radiation of electrons accelerated above the polar cap by a parallel electric field, derived assuming space-charge-limited flow (free emission of particles from the neutron star surface). A fraction of the photons undergo magnetic pair attenuation, producing a first-generation pair spectrum which then radiates synchrotron photons that produce further generations of pairs. The total cascade multiplicity (average number of pairs spawned by each primary electron) is a strong function of pulsar period $P$ and surface magnetic field strength $B_0$, so that many pulsars with low $B$-fields and long periods produce either few or no pairs, leading to a pair death line. 

\begin{figure}
\begin{center}
\includegraphics*[width=13cm,angle=0]{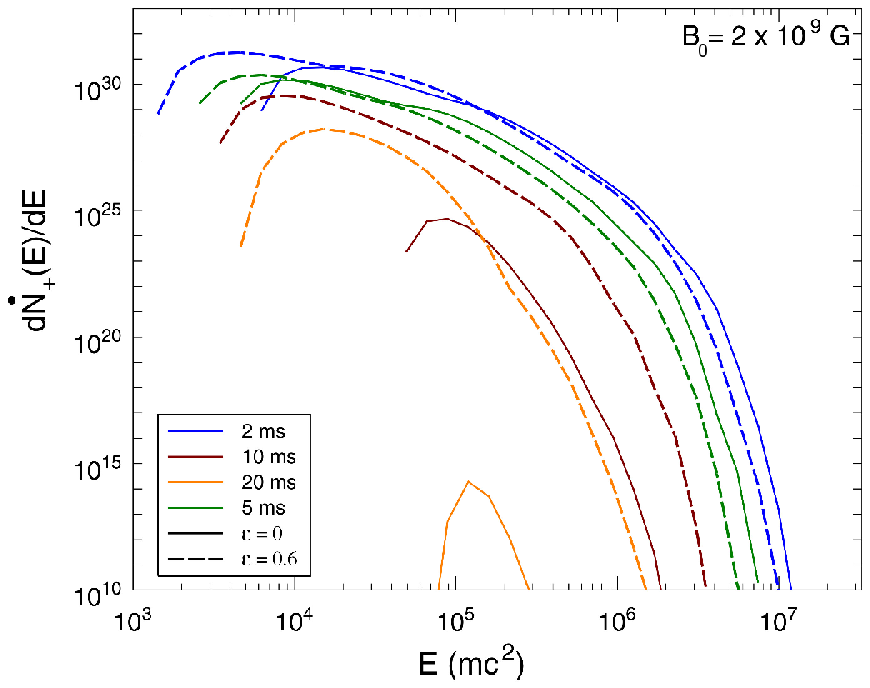}
\end{center}
\caption{Sample electron-positron pair spectra (number of pairs per second and energy) calculated for different periods and offset values ($P$ and $\varepsilon$), as indicated in the legend, and for a fixed value of $B_8 = 20$, i.e., $B_0 = 2\times10^9$~G. From \citet{Harding11}.}\label{fig:pair}
\end{figure}

The sweepback of magnetic field lines near the light cylinder (where the corotation speed equals the speed of light) as well as non-symmetric currents within the neutron star may cause the magnetic polar caps to be offset from the dipole axis. This may in turn lead to enhanced local electric fields, boosting pair formation, even for pulsars beyond the pair death line for pure dipole fields. We include this effect by using an offset-dipole magnetic field and find pair spectra (Figure~\ref{fig:pair}) that are quite hard, characterised by $P$, $\dot{P}$ (or equivalently, $B_0$), and offset parameter $\varepsilon$ \citep{Harding11}. From our first-principles cascade simulation, we find that about $\sim1$\% of $L_{\rm sd}$ is tapped to generate the pairs.

We used a grid in $P$ and $B_0$ 
and interpolated the spectra to find an adequate injection spectrum for each source in the present-day MSP population (Section~\ref{sec:pop}). We used an inclination angle of $\alpha=45^\circ$, mass $M = 2.15M_\odot$, radius $R = 9.9$~km, and moment of inertia $I = 1.56\times10^{45}$~g\,cm$^2$ for all MSPs\footnote{We adopted an equation of state with larger $M$ here (and associated smaller $I$) compared to that used in the population code (Section~\ref{sec:pop}), since some MSPs have measured masses $M\sim 2 M_\odot$ \citep{Demorest10}, and this enhances pair multiplicity. However, this discrepancy is removed by considering a large range of $\varepsilon$, since the latter simulates a large range of pair multiplicities that would correspond to different equations of state, and thus values of $M$.}. We used dipole offsets of $\varepsilon = (0.0,0.2,0.6)$ and set $\phi_0=\pi/2$ (this parameter controls the direction of offset of the polar cap).

Since MSPs are not surrounded by nebulae that can trap the pairs and degrade their energy before escape, we can assume that the pair spectra emerging from the MSPs are good representations of the intrinsic source spectra.  A number of MSPs are in binary systems and a subset of these, the black widows and redbacks, may contain strong interbinary shocks that can further accelerate the pairs. We will neglect the contribution from these sources here, but plan to treat this potentially important component in a future paper \citep{Venter15}.

\section{Interstellar Radiation Field (ISRF)}
\label{sec:ISRF}
In order to calculate IC losses suffered by leptons propagating through our Galaxy, one needs to know the spectral and spatial properties of Galactic ``background photons'' or ISRF. Optical stellar photons as well as infrared (IR) photons that result from scattering, absorption, and re-emission of the stellar photons by dust in the interstellar medium contribute to this field \citep{Porter08}. 

\begin{figure}
\begin{center}
\includegraphics*[width=13cm,angle=0]{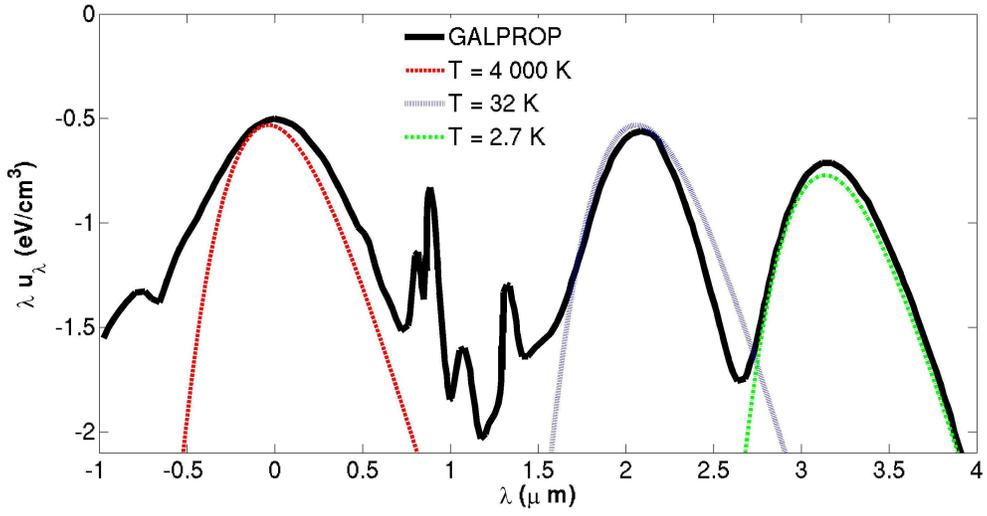}
\end{center}
\caption{The ISRF as calculated by \citet{Porter08} for the Galactic Plane, at a distance of 8~kpc from the Galactic Centre (thick black line), as well as three blackbody contributions which we use to approximate the ISRF (red dashed line: optical component with $T = 4~000$~K; blue dotted line: IR component with $T = 32$~K; and green dashed-dotted line: CMB with $T = 2.7$~K). Our calculations make use of the average energy density of each component, and not of the temperatures \textit{per se}.}\label{fig:ISRF2}
\end{figure}

The GALPROP code \citep{Strong98} includes a detailed ISRF model that incorporates a stellar population model, dust grain abundance and size distribution, as well as the absorption and scattering efficiencies of the latter (see also \citealt{Moskalenko06,Porter06,Porter08}). We find that the ISRF is adequately approximated by three blackbody components (optical, IR, and cosmic microwave background or CMB) as shown in Figure~\ref{fig:ISRF2} (for the Galactic Plane). We will use average photon energy densities to calculate the IC loss rate (see Eq.~[\ref{eq:transport}]). One can distinguish between two main spatial regions: the Galactic Plane and the Galactic Halo. For the Plane, we use $U_{\rm optical} = U_{\rm IR} = 0.4$~eV\,cm$^{-3}$, and $U_{\rm CMB} = 0.23$~eV\,cm$^{-3}$ \citep{Ruppel10}, while for the Halo, we use $U_{\rm optical} = 0.8$~eV\,cm$^{-3}$, $U_{\rm IR} = 0.05$~eV\,cm$^{-3}$, and $U_{\rm CMB} = 0.23$~eV\,cm$^{-3}$ \citep{Blies12}.

\section{The Galactic Magnetic Field}
\label{sec:Bfield}
We are interested in an average field strength that would determine SR loss rates (Eq.~[\ref{eq:SR}]), and not so much in the overall Galactic field structure (which is still under debate). The total field has been estimated to be around 6~$\mu$G, averaged over a distance of 1~kpc around the Sun, using radio SR measurements and equipartition arguments. This number increases to $\sim10\,\mu$G closer to the inner Galaxy \citep[see][and references therein]{Beck09}. \citet{Han06} used a combination of dispersion and rotation measures of over 500 pulsars and found that the regular $B$-field component\footnote{The Galactic $B$-field is typically divided into three components: regular, random, and anisotropic \citep{Beck09}.} decreases from $\sim6\,\mu$G near a Galactocentric distance of $2$~kpc to $\sim1\,\mu$G near $9$~kpc; the value is $\sim2\,\mu$G near the Sun \citep[see Figure~11 of][]{Han06}. Furthermore, the mean regular field as function of latitude is inferred to vary between $\sim-5\,\mu$G and $\sim5\,\mu$G \citep{Han06}. \citet{Orlando13} inferred values of $\sim2\,\mu$G, $\sim5\,\mu$G, and $\sim2\,\mu$G for the local regular, random, and anisotropic field components in the Disc via Galactic SR modelling. It is, however, important to note that the average total field decreases when taking into account its rapid decay with height above the Plane. \citet[][hereafter D10]{D10} therefore argue that the relevant $B$-field (the square root of the sum of averages of the squares of all field components) to be used in SR calculations may reasonably be $1 - 3\,\mu$G.

\section{Total Leptonic Energy Loss Rate}
\label{sec:losses}
The SR loss rate is given by
  \begin{equation}
  \dot{E}_{\rm SR} = \frac{4\sigma_{\rm T}cU_BE^2}{3\left(m_ec^2\right)^2},\label{eq:SR}
  \end{equation}
with $\sigma_{\rm T}$ the Thomson cross section, $E$ the lepton energy, and the magnetic energy density
  \begin{equation}
  U_B = \frac{B^2}{8\pi} = 0.098b_2^2~{\rm eV\,cm}^{-3}
  \end{equation}
for $b_2 = B/(2.0~\mu$G). The general expression for the IC loss rate (for target photons of energy density $U_i$, for the $i^{\rm th}$ blackbody component) may be approximated as \citep[for details, see the appendix of][]{Ruppel10}
  \begin{equation}
\dot{E}_{\rm IC} = \frac{4\sigma_{\rm T}cU_iE^2}{3\left(m_ec^2\right)^2}\frac{\gamma_K^2}{\gamma_K^2 + \gamma^2},
  \end{equation}
with the critical Klein-Nishina Lorentz factor defined as
  \begin{equation}
\gamma_K \equiv \frac{3\sqrt{5}}{8\pi}\frac{m_ec^2}{k_BT} \approx \frac{0.27m_ec^2}{k_BT},
  \end{equation}
and $m_e$ the electron mass. If $\gamma\ll \gamma_K$, we recover the well-known expression for the Thomson limit \citep{Blumenthal70}
  \begin{equation}
\dot{E}_{\rm IC} = \frac{4\sigma_{\rm T}cU_iE^2}{3\left(m_ec^2\right)^2}.
\end{equation}
The Klein-Nishina correction is only necessary for optical photons, where $\gamma_K\sim10^5$. The IC loss rate for particles with Lorentz factors above $\gamma_K$ is severely suppressed. For simplicity, one may choose to neglect this correction, which implies an overestimation of the IC loss rate for particles at the highest energies. This, however, allows the introduction of an effective $B$-field, $B_{\rm eff} = qB$, which simulates the effect of adding the IC and SR losses \citep{Blies12}:
  \begin{eqnarray}
\dot{E} &=& \dot{E}_{\rm SR} + \dot{E}_{\rm IC} = \frac{4\sigma_{\rm T}cE^2}{3\left(m_ec^2\right)^2}\left[U_B + \sum_iU_i \right]\nonumber\\
& = &\frac{4\sigma_{\rm T}cE^2}{3\left(m_ec^2\right)^2}U_{\rm eff} = b_0E^2,\label{eq:loss_b}\\
  U_{\rm eff} & = & U_B(1+f) = \frac{B_{\rm eff}^2}{8\pi} = q^2\frac{B^2}{8\pi},\\
  f & = & \frac{\sum_iU_i}{U_B},\\
  q & = & \sqrt{1+f},\\
  b_0 & = & \frac{4c}{9}\left(\frac{e}{m_ec^2}\right)^4B_{\rm eff}^2 = 1.58\times10^{-15}\left(\frac{B_{\rm eff}}{1~\mu{\rm G}}\right)^2.
  \end{eqnarray}
If one considers typical constant average values of $B=2-3~\mu$G for the Galactic Plane and $B=1\,\mu$G
for the Galactic Halo, one ends up with $B_{\rm eff}\approx7\,\mu$G in both cases. This is because the increase in $U_{\rm optical}$ is coupled with the decrease in $B$ as one moves from the Galactic Plane to the Halo. However, to account for the fact that the Thomson loss rate overestimates the true IC loss rate, we will rather use values of $B_{\rm eff} = 2~\mu$G and $B_{\rm eff} = 5~\mu$G, which should give a reasonable range for the predictions of cosmic-ray lepton flux from nearby MSPs.

\section{Transport Model}
\label{sec:transport}
We solve a Fokker-Planck-type transport equation
  \begin{equation}
\frac{\partial n_{\rm e}}{\partial t}=\mathbf{\nabla}\cdot\left({\cal K}\cdot\mathbf{\nabla} n_{\rm e} \right)-\frac{\partial}{\partial E}\left(\dot E n_{\rm e}\right)+S,\label{eq:transport}
  \end{equation}
with $n_{\rm e}$ the lepton density (per energy interval) and $E$ the lepton energy. Also, ${\cal K}$ denotes the diffusion tensor and $\dot E$ energy losses, while $S$ is the source term. We consider a steady-state case, as well as spatially-independent diffusion (so that ${\cal K}$ becomes only a function of energy $\kappa(E)$), assuming a power-law dependence on energy:
  \begin{equation}
  \kappa(E) = \kappa_0\left(\frac{E}{E_{\rm norm}}\right)^\alpha_{\rm D},
  \end{equation}
and we assume typical values of $\alpha_{\rm D} = 0.6$, $E_{\rm norm} = 1$~GeV, and $\kappa_0 = 0.1~{\rm kpc}^2{\rm Myr}^{-1} \approx 3\times10^{28}$~cm$^2$s$^{-1}$ \citep[e.g.,][]{Moskalenko98}. The source term refers to the pair spectra $Q_i(P,B_0,\varepsilon,E)$ of $N\sim4\times10^4$ Galactic MSP sources (i.e., for the $i^{\rm th}$ pulsar in our population, we assign a pair spectrum $Q_i(P,B_0,\varepsilon,E)$, as calculated in Section~\ref{sec:pairs} for the corresponding simulated values of $P$, $B_0$, and $\varepsilon$; the properties of this population have been calculated in Section~\ref{sec:pop}):
  \begin{equation}
    S = \sum_i^NQ_i(E)\delta(\mathbf{r} - \mathbf{r}_{0,i}).
  \end{equation}
Here, $\mathbf{r}_{0,i}$ are the source positions. For an infinite system, Eq.~(\ref{eq:transport}) is solved by the following Green's function \citep{D10,Blies12}:
\begin{equation}
G(\mathbf{r},\mathbf{r}_0,E,E_0) = \frac{\Theta(E_0 - E)}{\dot{E}\left(\pi\lambda\right)^{3/2}}\exp\left(-\frac{|\mathbf{r} - \mathbf{r}_0|^2}{\lambda}\right) 
\end{equation}
with $E_0$ the particle energy at the source, and the square of the propagation scale characterised by 
\begin{eqnarray}
\lambda(E,E_0) & \equiv & 4\int_{E}^{E_0}\frac{\kappa(E^\prime)}{b(E^\prime)}\,dE,^\prime\\
 & = & \lambda_0\left[\frac{1}{E_0}\left(\frac{E_0}{E_{\rm norm}}\right)^{\alpha_{\rm D}} - \frac{1}{E}\left(\frac{E}{E_{\rm norm}}\right)^{\alpha_{\rm D}}\right],\\
\lambda_0 & = & \frac{4\kappa_0}{\left(\alpha_{\rm D}-1\right)b_0},
\end{eqnarray}
and $\Theta(E_0 - E) $ the Heaviside function. The latter is used to ensure that $\lambda>0$.
The lepton flux may then be found using
\begin{equation}
\phi_e(\mathbf{r},E) = \frac{c}{4\pi}\int\!\!\!\!\int\!\!\!\!\int\!\!\!\!\int G(\mathbf{r},\mathbf{r}_0,E,E_0)S\,dE_0d^3r_0.
\end{equation}
While the finite boundary of the Galactic Halo should impact the solution, this effect is not too large for GeV leptons, for which the propagation scale is only a few kpc (D10), and we neglect it here for simplicity. Our results will indicate that our predicted MSP contribution becomes significant above $\sim10$~GeV, so that the effect of solar modulation may safely be neglected \citep{Strauss14}.

\section{Results}
\label{sec:results}
\begin{figure}[t]
\begin{center}
\includegraphics*[width=13cm,angle=0]{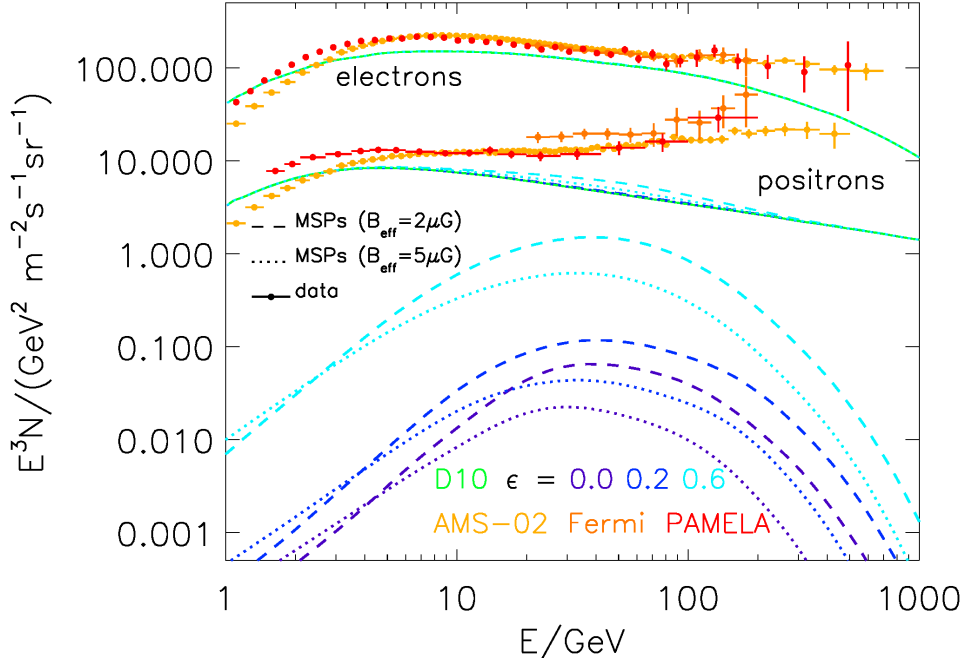}
\end{center}
\caption{MSP contribution to the cosmic-ray lepton spectra at Earth. Electron spectra appear at the top, and positron spectra lower. The solid green lines indicate the ``background'' component of primary electrons from distant supernova remnants plus secondary electrons (upper solid line), as well as a component of secondary positrons (lower solid line) as predicted by D10. The hump-like primary MSP spectra (only positrons are shown, but the electrons are assumed to have identical spectra) appear at the bottom of the plot. From bottom to top (purple, blue, and cyan), these spectra are for $\varepsilon = 0.0, 0.2,$ and $0.6$. The addition of the various MSP components to the background flux of D10 is indicated using the same colour convention. Dashed lines are for $B_{\rm eff}=2\,\mu$G and dotted lines for $B_{\rm eff}=5\,\mu$G. Also shown are data from \textit{PAMELA} \citep[red;][]{Adriani13}, \textit{Fermi} \citep[orange;][]{Ackermann12}, and \textit{AMS$-$02}  \citep[yellow;][]{Aguilar14}, accessed via the website http:$/\!/$lpsc.in2p3.fr$/$cosmic-rays-db \citep{Maurin14}.}\label{fig:LIS}
\end{figure}

Figure~\ref{fig:LIS} indicates the ``background'' electron and positron fluxes predicted by D10 (we use their secondary positron flux as well as the sum of their secondary electron flux and primary electron flux originating in distant supernova remnants, as indicated in their Figure~14), as well as data from \textit{Fermi} \citep{Ackermann12}, \textit{PAMELA} \citep{Adriani13}, and \textit{AMS$-$02} \citep{Aguilar14}. We furthermore indicate primary electrons and positrons from MSPs for dipole offsets of $\varepsilon = (0.0, 0.2, 0.6)$ and Galactic $B$-fields of $B_{\rm eff} = 2\,\mu$G and $B_{\rm eff} = 5\,\mu$G. The various curves are distinguished in the Figure caption. 

\begin{figure}
\begin{center}
\includegraphics*[width=13cm,angle=0]{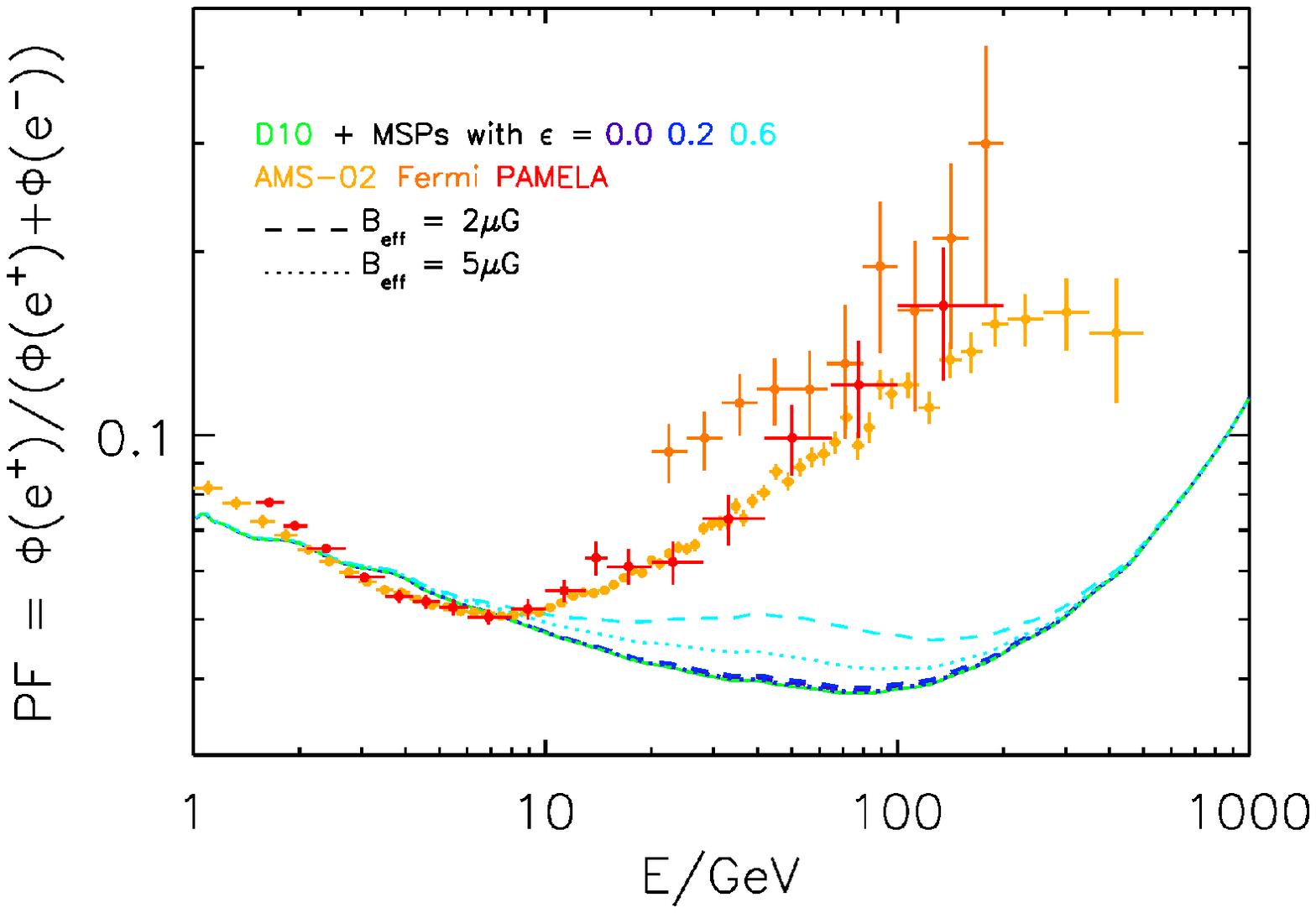}
\end{center}
\caption{Measured \citep{Ackermann12,Adriani13,Accardo14} and predicted PF, including ``background'' contributions from D10 indicated by a solid green line, and the MSP contribution from this work. We use the same colour scheme and line styles as in Figure~\ref{fig:LIS} to distinguish between values of $\varepsilon$ and $B_{\rm eff}$ (see legend).}\label{fig:PF}
\end{figure}

Figure~\ref{fig:PF} shows the measured PF \citep[e.g.,][]{Accardo14} as well as the D10 and (D10 + MSP) contributions. The largest contribution is found in the case of $\varepsilon=0.6$ and $B=2\,\mu$G. The contribution is lower for $\varepsilon=0.6$ and $B_{\rm eff}=5\,\mu$G, and negligible for $\varepsilon=0.0$ and $\varepsilon=0.2$. 

\section{Conclusion}
\label{sec:concl}
The predicted MSP contribution increases for non-zero values of polar cap offset parameter $\varepsilon$. This is expected, since a larger value for the offset of the surface $B$-field with respect to the non-perturbed magnetic axis leads to an increase in the acceleration potential for some regions in azimuthal phase (and a decrease in others). This in turn results in an enhancement in both the number of particles (since the multiplicity will be higher) as well as the maximum particle source energy. We find that the MSPs make only a modest contribution to the local cosmic-ray flux at a few tens of GeV, after which this spectral component cuts off. 

The effect of different Galactic $B$-fields is also shown. Increasing $B_{\rm eff}$ leads to an increased energy loss rate $\dot{E}_{\rm SR}(E)\propto B_{\rm eff}^2$ and a decrease in particle flux. We will address the effect of including Klein-Nishina corrections to the total loss rate in a future work \citep{Venter15}. The result should however be in a similar range, while the spectral shape may change somewhat, since the high-energy particles will suffer smaller loss rates and hence have longer survival times.

Although the PF is somewhat enhanced above 10~GeV (also depending on the ``background'' model for the secondary positron flux), our added MSP component fails to reproduce the high-energy rise for the parameters considered. This implies alternative sources of primary positrons such as young, nearby pulsars or supernova remnants \citep[e.g.,][]{DiMauro14}. 

\citet{Kisaka12} suggested that pair-starved MSPs may be responsible for a large peak in the total electron spectrum at $10-100$~TeV, and that non-pair-starved MSPs with multiplicities of $\sim2~000$ may contribute significantly (near 100\%) to the PF above 10~GeV. There are, however, a number of differences in our respective approaches. \citet{Kisaka12} use fixed values for $P$ and $B_0$ for all members in their population. They furthermore assumed energy equipartition between the particles and the $B$-field, which seems to imply a conversion efficiency (from spin-down luminosity to particle power) of $\eta\sim 50\%$, while we find $\eta\sim1\%$ from our simulations. Their diffusion coefficient is slightly larger than ours, and they assume a lower average Galactic $B$-field ($B = 1\,\mu$G) as well as including a Klein-Nishina correction for their IC losses. Finally, they integrate the injected spectra over the age of the MSPs 
while we follow a steady-state approach. Most if not all of these differences should lead to an enhanced particle flux in their case.

Our population synthesis uses radio survey sensitivity and \textit{Fermi} three-year point source sensitivity maps, normalising to the number of detected radio MSPs from those surveys only, and to the detected $\gamma$-ray MSPs in the 2nd \textit{Fermi} pulsar catalogue, all of which are radio-loud. The MSPs discovered in radio followup observations of unidentified \textit{Fermi} sources are included in our simulated population of MSPs not detected by surveys but detected by \textit{Fermi} as point sources. However, it is possible that there is a contribution to the cosmic-ray flux from MSPs in binary systems (e.g., black widows or redbacks) very close to Earth that, due to difficulty of radio detection in these eclipsing and/or obscured systems, have not been identified. These systems may further accelerate the electron-positron pairs in the strong interbinary shocks, and we will investigate this potentially important contribution in a future paper. We will lastly also consider the effect of different assumptions for the magnitude of the spatial diffusion coefficient \citep{Venter15}.

\clearpage

\clearpage

\clearpage

\clearpage


\begin{thebibliography}{}
\bibitem[Abdo et al.(2013)]{Abdo13} Abdo, A.A. et al., The second \textit{Fermi} Large Area Telescope catalog of gamma-ray pulsars, ApJS, 208, 17-75, 2013.
\bibitem[Accardo et al.(2014)]{Accardo14} Accardo, L. et al., High statistics measurement of the positron fraction in primary cosmic rays of 0.5$-$500 GeV with the Alpha Magnetic Spectrometer on the International Space Station, Phys.\ Rev.\ Lett., 113, 121101, 2014.
\bibitem[Ackermann et al.(2012)]{Ackermann12} Ackermann, M. et al., Measurement of separate cosmic-ray electron and positron spectra with the \textit{Fermi} Large Area Telescope, Phys.\ Rev.\ Lett., 108, 011103, 2012.
\bibitem[Adriani et al.(2009)]{Adriani09} Adriani, O. et al., An anomalous positron abundance in cosmic rays with energies 1.5$-$100~GeV, Nature, 458, 607-609, 2009.
\bibitem[Adriani et al.(2013)]{Adriani13} Adriani, O. et al., Cosmic-ray positron energy spectrum measured by \textit{PAMELA}, Phys.\ Rev.\ Lett., 111, 081102, 2013.
\bibitem[Aguilar et al.(2013)]{Aguilar13} Aguilar, M. et al., First result from the Alpha Magnetic Spectrometer on the International Space Station: precision measurement of the positron fraction in primary cosmic rays of 0.5$-$350~GeV, Phys.\ Rev.\ Lett., 110, 141102, 2013.
\bibitem[Aguilar et al.(2014)]{Aguilar14} Aguilar, M. et al., Electron and positron fluxes in primary cosmic rays measured with the Alpha Magnetic Spectrometer on the International Space Station, Phys.\ Rev.\ Lett., 113, 121102, 2014.
\bibitem[Beck(2009)]{Beck09} Beck, R., Galactic and extragalactic magnetic fields - a concise review, Astrophys.\ Space Sci.\ Trans., 5, 43-47, 2009.
\bibitem[Blasi(2009)]{Blasi09} Blasi, P., Origin of the positron excess in cosmic rays, Phys.\ Rev.\ Lett., 103, 051104, 2009.
\bibitem[Blies \& Schlickeiser(2012)]{Blies12} Blies, P., \& Schlickeiser, R., The influence of Klein-Nishina steps on the spatial diffusion of Galactic cosmic-ray electrons, ApJ, 751, 71-77, 2012.
\bibitem[Blum et al.(2013)]{Blum13} Blum, K., Katz, B., \& Waxman, E., AMS-02 Results Support the Secondary Origin of Cosmic Ray Positrons, PRL, 111, 211101, 2013.
\bibitem[Blumenthal \& Gould(1970)]{Blumenthal70} Blumenthal, G.R., \& Gould, R.J., Bremsstrahlung, synchrotron radiation, and Compton scattering of high-energy electrons traversing dilute gases, Rev.\ Mod.\ Phys., 42, 237-271, 1970.
\bibitem[B\"usching et al.(2008a)]{Buesching08} B{\"u}sching, I., Venter, C., \& de Jager, O.C., Contributions from nearby pulsars to the local cosmic ray electron spectrum, Adv.\ Space Res., 42, 497-503, 2008a.
\bibitem[B\"usching et al.(2008b)]{Buesching08b} B{\"u}sching, I., de Jager, O.C., Potgieter, M.S., \& Venter, C., A cosmic-ray positron anisotropy due to two middle-aged, nearby pulsars?, ApJ, 678, L39-L42, 2008b.
\bibitem[Contopoulos et al.(2014)]{Contopoulos14} Contopoulos, I., Kalapotharakos, C., \& Kazanas, D., A new standard pulsar magnetosphere, ApJ, 781, 46-50, 2014.
\bibitem[Cowsik \& Burch(2010)]{Cowsik10} Cowsik, R., \& Burch, B., Positron fraction in cosmic rays and models of cosmic-ray propagation, Phys.\ Rev.\ D, 82, 023009, 2010.
\bibitem[Daugherty \& Harding(1982)]{DH82} Daugherty, J.K. \& Harding, A.K., Electromagnetic cascades in pulsars, ApJ, 252, 337-347, 1982.
\bibitem[Delahaye et al.(2010)]{D10} Delahaye, T., Lavalle, J., Lineros, R., Donato, F., \& Fornengo, N., Galactic electrons and positrons at the earth: new estimate of the primary and secondary fluxes, A\&A, 524, A51, 2010 (D10).
\bibitem[Demorest et al.(2010)]{Demorest10} Demorest, P.B., Pennucci, T., Ransom, S.M., Roberts, M.S.E., \& Hessels, J.W.T., A two-solar-mass neutron star measured using Shapiro delay, Nature, 467, 1081-1083, 2010.
\bibitem[Di Mauro et al.(2014)]{DiMauro14} Di Mauro, M., Donato, F., Fornengo, N., Lineros, R., \& Vittino, A., Interpretation of AMS-02 electrons and positrons data, JCAP, 4, 6-39
\bibitem[Fan et al.(2010)]{Fan10} Fan, Y.-Z., Zhang, B., \& Chang, J., Electron/positron excesses in the cosmic ray spectrum and possible interpretations, Int.\ J.\ Mod.\ Phys.\ D, 19, 2011-2058, 2010.
\bibitem[Gonthier et al.(2015)]{Gonthier15} Gonthier, P.L. et al., Population synthesis of radio and gamma-ray millisecond pulsars from the Galactic disk, in prep.
\bibitem[Gupta \& Torres(2014)]{Gupta14} Gupta, N. \& Torres, D.F., p$\gamma$ interactions in Galactic jets as a plausible origin of the positron excess, MNRAS, 441, 3122-3126, 2014.
\bibitem[Han et al.(2006)]{Han06} Han, J.L., Manchester, R.N., Lyne, A.G., Qiao, G.J., \& van Straten, W., Pulsar rotation measures and the large-scale structure of the Galactic magnetic field, ApJ, 642, 868-881, 2006.
\bibitem[Harding \& Muslimov(2011)]{Harding11} Harding, A.K., \& Muslimov, A.G., Pulsar pair cascades in magnetic fields with offset polar caps, ApJ, 743, 181-196, 2011.
\bibitem[Hobbs et al.(2005)]{Hobbs05} Hobbs, G., Lorimer, D.R., Lyne, A.G., \& Kramer, M., A statistical study of 233 pulsar proper motions, MNRAS, 360, 974-992, 2005.
\bibitem[Kisaka \& Kawanaka(2012)]{Kisaka12} Kisaka, S., \& Kawanaka, N., TeV cosmic-ray electrons from millisecond pulsars, MNRAS, 421, 3543-3549, 2012.
\bibitem[Lamb \& Yu(2005)]{Lamb2005} Lamb, F., \& Yu, W., Spin rates and magnetic fields of millisecond pulsars,
in Binary Radio Pulsars, ASP Conf.\ Ser., ed.\ F.A. Rasio \& I.H.\ Stairs, 328, 299-310, 2005.
\bibitem[Li et al.(2012)]{Li12} Li, J., Spitkovsky, A., \& Tchekhovskoy, A., Resistive solutions for pulsar magnetospheres, ApJ, 746, 60-71, 2012.
\bibitem[Maurin et al.(2014)]{Maurin14} Maurin, D., Melot, F., \& Taillet, R., A database of charged cosmic rays, A\&A, 569, A32, 2014.
\bibitem[Moskalenko \& Strong(1998)]{Moskalenko98} Moskalenko, I.V., \& Strong, A.W., Production and propagation of cosmic-ray positrons and electrons, ApJ, 493, 694-707, 1998.
\bibitem[Moskalenko et al.(2006)]{Moskalenko06} Moskalenko, I.V., Porter, T.A., \& Strong, A.W., Attenuation of very high energy gamma rays by the Milky Way interstellar radiation field, ApJ, 640, L155-L158, 2006.
\bibitem[Moskalenko(2013)]{Moskalenko13} Moskalenko, I.V., Cosmic rays in the Milky Way and beyond, Nuclear Phys.\ B Proc.\ Suppl., 243, 85-91, 2013.
\bibitem[Orlando \& Strong(2013)]{Orlando13} Orlando, E. \& Strong, A., Galactic synchrotron emission with cosmic ray propagation models, MNRAS, 436, 2127-2142, 2013.
\bibitem[Paczy\'nski(1990)]{Pacz90} Paczy\'nski, B., A test of the galactic origin of gamma-ray bursts, ApJ, 348, 485-494, 1990.
\bibitem[Pierbattista et al.(2012)]{Pierbattista12} Pierbattista, M., Grenier, I.A., Harding, A.K., \& Gonthier, P.L., Constraining {$\gamma$}-ray pulsar gap models with a simulated pulsar population, A\&A, 545, A42, 2012.
\bibitem[Porter et al.(2006)]{Porter06} Porter, T.A., Moskalenko, I.V., \& Strong, A.W., Inverse Compton emission from Galactic supernova remnants: effect of the interstellar radiation field, ApJ, 648, L29-L32, 2006. 
\bibitem[Porter et al.(2008)]{Porter08} Porter, T.A., Moskalenko, I.V., Strong, A.W., Orlando, E., \& Bouchet, L., 
Inverse Compton origin of the hard X-ray and soft gamma-ray emission from the Galactic ridge, ApJ, 682, 400-407, 2008.
\bibitem[Porter et al.(2011)]{Porter11} Porter, T.A., Johnson, R.P., \& Graham, P.W., Dark matter searches with astroparticle data, Ann.\ Rev.\ Astron.\ Astrophys., 49, 155-194, 2011.
\bibitem[Serpico(2012)]{Serpico12} Serpico, P.D., Astrophysical models for the origin of the positron ``excess'', Astropart.\ Phys., 39, 2-11, 2012.
\bibitem[Schlickeiser \& Ruppel(2010)]{Ruppel10} Schlickeiser, R., \& Ruppel, J., Klein-Nishina steps in the energy spectrum of galactic cosmic-ray electrons, New J.\ Phys., 12, 033044, 2010.
\bibitem[Shaviv et al.(2009)]{Shaviv09} Shaviv, N.J., Nakar, E., \& Piran, T., Inhomogeneity in Cosmic Ray Sources as the Origin of the Electron Spectrum and the \textit{PAMELA} Anomaly, PRL, 103, 111302, 2009.
\bibitem[Spitkovsky(2006)]{Spitkovsky06} Spitkovsky, A., Time-dependent force-free pulsar magnetospheres: axisymmetric and oblique rotators, ApJ,648, L51-L54, 2006.
\bibitem[Story et al.(2007)]{Story07} Story, S.A., Gonthier, P.L., \& Harding, A.K., Population synthesis of radio and $\gamma$-ray millisecond pulsars from the Galactic disk, ApJ, 671, 713-726, 2007.
\bibitem[Strauss \& Potgieter(2014)]{Strauss14} Strauss, R.D., \& {Potgieter}, M.S., Where does the heliospheric modulation of galactic cosmic rays start?, Adv.\ Space Res., 53, 1015, 2014
\bibitem[Strong \& Moskalenko(1998)]{Strong98} Strong, A.W., \&  Moskalenko, I.V., Propagation of cosmic-ray nucleons in the Galaxy, ApJ, 509, 212-228, 1998.
\bibitem[Sturrock(1971)]{Sturrock71} Sturrock, P.A., A model of pulsars, ApJ, 164, 529-556, 1971.
\bibitem[Venter et al.(2009)]{Venter09} Venter, C., Harding, A.K., \& Guillemot, L., Probing millisecond pulsar emission geometry using light curves from the \textit{Fermi}/Large Area Telescope, ApJ, 707, 800-822, 2009.
\bibitem[Venter et al.(submitted for publication)]{Venter15} Venter, C., Kopp, A., Gonthier, P.L., Harding, A.K., \& B{\"u}sching, I., Cosmic-ray positrons from millisecond pulsars, submitted to ApJ.
\bibitem[Zhang \& Cheng(2001)]{Zhang01} Zhang, L., \& Cheng, K.S., Cosmic-ray positrons from mature gamma-ray pulsars, A\&A, 368, 1063-1070, 2001. 
\end{thebibliography}
\end{document}